\documentclass{rnti}
\pdfoutput=1
\usepackage{amsmath,amssymb}
\usepackage{theorem}
\usepackage{graphicx}

\usepackage{ifpdf}
\usepackage{color}
\ifpdf
\else

\fi

\newtheorem{defn}{D\'{e}finition}

\begin{document}

\thispagestyle{empty}

\fancyhead{} \fancyfoot{} \fancyhead[LE]{Espaces de repr\'{e}sentation
multidimensionnels d\'{e}di\'{e}s \`{a} la visualisation} \fancyhead[RO]{Ben Messaoud {\it
et al.}} \fancyfoot[LE,RO]{RNTI - E - ~}

\begin{center}
{\Large\bf Une approche de construction d'espaces \\ de repr\'{e}sentation
multidimensionnels d\'{e}di\'{e}s \`{a} la visualisation}
~\\~\\ Riadh Ben Messaoud, Kamel Aouiche, C\'{e}cile Favre\\ ~\\
Laboratoire ERIC, Universit\'{e} Lumi\`{e}re Lyon 2\\ 5 avenue Pierre
Mend\`{e}s-France\\69676 Bron Cedex\\ \{rbenmessaoud $|$ kaouiche $|$
cfavre\}@eric.univ-lyon2.fr
\end{center}

\begin{abstract}
Dans un syst\`{e}me d\'{e}cisionnel, la composante visuelle est importante
pour l'analyse en ligne OLAP. Dans cet article, nous proposons une
nouvelle approche qui permet d'apporter une solution au probl\`{e}me
de visualisation des donn\'{e}es engendr\'{e} par l'\'{e}parsit\'{e}. En se basant
sur les r\'{e}sultats d'une analyse des correspondances multiples
(ACM), nous tentons d'att\'{e}nuer l'effet n\'{e}gatif de l'\'{e}parsit\'{e} en
organisant diff\'{e}remment les cellules d'un cube de donn\'{e}es. Notre
m\'{e}thode ne cherche pas \`{a} r\'{e}duire l'\'{e}parsit\'{e} mais plut\^{o}t \`{a}
construire un espace de repr\'{e}sentation se pr\^{e}tant mieux \`{a}
l'analyse et dans lequel les faits du cube sont regroup\'{e}s. Pour
\'{e}valuer l'apport de cette nouvelle repr\'{e}sentation des donn\'{e}es,
nous proposons un indice d'homog\'{e}n\'{e}it\'{e} bas\'{e} sur le voisinage
g\'{e}om\'{e}trique des cellules d'un cube. Les diff\'{e}rents tests men\'{e}s
nous ont montr\'{e} l'efficacit\'{e} de notre m\'{e}thode.

\noindent \textbf{Mots-cl\'{e}s :} ACM, arrangement, cube de donn\'{e}es,
\'{e}parsit\'{e} d'un cube, espace de repr\'{e}sentation, indice
d'homog\'{e}n\'{e}it\'{e}, OLAP, visualisation, voisinage.
\end{abstract}

\section{Introduction}\label{introduction}

Dans un contexte concurrentiel d\'{e}velopp\'{e}, les entreprises telles que les
banques\footnote{Nous remercions Michel Rougi\'{e}, repr\'{e}sentant du Cr\'{e}dit
Lyonnais, pour les donn\'{e}es fournies afin de valider ce travail.} doivent
aujourd'hui \^{e}tre capables de prendre des d\'{e}cisions pertinentes, de fa\c{c}on
r\'{e}active. La mise en place d'un processus d\'{e}cisionnel est alors n\'{e}cessaire pour
g\'{e}rer une masse de donn\'{e}es de plus en plus cons\'{e}quente. Le stockage et la
centralisation de ces donn\'{e}es dans un entrep\^{o}t constitue un support efficace
pour l'analyse de ces derni\`{e}res. En effet, \`{a} partir d'un entrep\^{o}t de donn\'{e}es,
on dispose d'outils permettant de construire des contextes d'analyse
multidimensionnels cibl\'{e}s, appel\'{e}s commun\'{e}ment cubes de donn\'{e}es. Ces cubes de
donn\'{e}es r\'{e}pondent \`{a} des besoins d'analyse pr\'{e}d\'{e}finis en amont.

L'analyse en ligne OLAP (On Line Analytical Processing) est un outil bas\'{e} sur
la visualisation permettant la navigation, l'exploration dans ces cubes de
donn\'{e}es. L'objectif est d'observer des faits, \`{a} travers une ou plusieurs
mesures, en fonction de diff\'{e}rentes dimensions. Il s'agit par exemple
d'observer les niveaux de ventes en fonction des produits, des p\'{e}rim\`{e}tres
commerciaux (localisations g\'{e}ographiques) et de la p\'{e}riode d'achat.

De cette visualisation d\'{e}pend la qualit\'{e} d'exploitation des donn\'{e}es. Or,
diff\'{e}rents facteurs peuvent d\'{e}grader cette visualisation. D'une part, la
repr\'{e}sentation multidimensionnelle engendre une \'{e}parsit\'{e}, puisqu'\`{a}
l'intersection de diff\'{e}rentes modalit\'{e}s de dimensions, il n'existe pas
forc\'{e}ment de faits correspondants. Cette \'{e}parsit\'{e} peut \^{e}tre accentu\'{e}e par la
consid\'{e}ration d'un grand nombre de dimensions (forte dimensionnalit\'{e}) et/ou
d'un grand nombre de modalit\'{e}s dans chacune des dimensions. D'autre part, les
modalit\'{e}s des dimensions sont g\'{e}n\'{e}ralement repr\'{e}sent\'{e}es selon un ordre
pr\'{e}-\'{e}tabli (ordre naturel) : ordre chronologique pour les dates, alphab\'{e}tique
pour les libell\'{e}s. Dans la plupart des cas, cet ordre entra\^{\i}ne une distribution
al\'{e}atoire des points repr\'{e}sentant les faits observ\'{e}s (les cellules pleines)
dans l'espace des dimensions.

Dans cet article, nous proposons d'am\'{e}liorer la visualisation des donn\'{e}es dans
les cubes. Nous ne diminuons pas l'\'{e}parsit\'{e} du cube comme dans~\cite{NNT03},
mais \`{a} att\'{e}nuer son effet n\'{e}gatif sur la visualisation, en regroupant les
cellules pleines. Pour ce faire, nous proposons d'arranger l'ordre des
modalit\'{e}s \'{e}tant donn\'{e} que l'ordre initial n'engendre pas forc\'{e}ment une bonne
visualisation. Cet arrangement tient compte des corr\'{e}lations existant entre les
faits pr\'{e}sents dans l'espace de repr\'{e}sentation d'un cube de donn\'{e}es. Les
corr\'{e}lations sont fournies par le r\'{e}sultat d'une analyse des correspondances
multiples (ACM) appliqu\'{e}e sur les faits du cube.

Ce travail s'inscrit dans une approche g\'{e}n\'{e}rale de couplage entre fouille de
donn\'{e}es et analyse en ligne. Dans~\cite{BRB05}, une r\'{e}flexion sur l'usage de
l'analyse factorielle dans un contexte OLAP a \'{e}t\'{e} amorc\'{e}e. \`A pr\'{e}sent, nous
exploitons l'ACM comme un outil d'aide \`{a} la construction de cubes de donn\'{e}es
ayant de meilleures caract\'{e}ristiques pour la visualisation. En effet, l'ACM
construit des axes factoriels qui offrent de meilleurs points de vue du nuage
de points des individus.

L'article est organis\'{e} comme suit. Dans la section~\ref{contexte}, nous
repositionnons plus en d\'{e}tail le contexte et les motivations de notre travail.
Nous d\'{e}taillons les diff\'{e}rentes \'{e}tapes de notre approche dans la
section~\ref{formalisation}. Nous pr\'{e}sentons dans la section~\ref{etude} une
\'{e}tude de cas sur un jeu de donn\'{e}es bancaires. Dans la section~\ref{discussion},
nous donnons un aper\c{c}u des travaux connexes au n\^{o}tre. Enfin, dans la
section~\ref{conclusion}, nous dressons une conclusion et proposons des
perspectives de recherche.

\section{Contexte et motivations}\label{contexte}

Dans un syst\`{e}me d\'{e}cisionnel, les donn\'{e}es sont organis\'{e}es selon un mod\`{e}le, en
{\it ``\'{e}toile''} ou en {\it ``flocon~de~neige''}, d\'{e}di\'{e} \`{a} l'analyse et
traduisant un contexte d'\'{e}tude cibl\'{e}~\cite{Inm96,Kim96}. Autour d'une table de
faits centrale contenant une ou plusieurs mesures \`{a} observer, existent
plusieurs tables de dimensions comprenant des descripteurs. Une dimension peut
comporter plusieurs hi\'{e}rarchies impliquant diff\'{e}rents niveaux de granularit\'{e}s
possibles dans la description de chaque fait. Cette organisation est
particuli\`{e}rement adapt\'{e}e pour cr\'{e}er des structures multidimensionnelles,
appel\'{e}es {\it ``cubes''} de donn\'{e}es, destin\'{e}es \`{a} l'analyse OLAP. Dans un cube
de donn\'{e}es, un fait est ainsi identifi\'{e} par un ensemble de modalit\'{e}s prises par
les diff\'{e}rentes dimensions. Le fait est observ\'{e} par une ou plusieurs mesures
ayant des propri\'{e}t\'{e}s d'additivit\'{e} plus ou moins fortes.

La vocation de l'OLAP est de fournir \`{a} l'utilisateur un outil visuel pour
consulter, explorer et naviguer dans les donn\'{e}es d'un cube afin d'y d\'{e}couvrir
rapidement et facilement des informations pertinentes. Toutefois, dans le cas
de donn\'{e}es volumineuses, telles que les donn\'{e}es bancaires consid\'{e}r\'{e}es dans
notre \'{e}tude, l'analyse en ligne n'est pas une t\^{a}che facile pour l'utilisateur.
En effet, un cube \`{a} forte dimensionnalit\'{e} comportant un grand nombre de
modalit\'{e}s, pr\'{e}sente souvent une structure \'{e}parse difficile \`{a} exploiter
visuellement. De plus, l'\'{e}parsit\'{e}, souvent r\'{e}partie de fa\c{c}on al\'{e}atoire dans le
cube, alt\`{e}re davantage la qualit\'{e} de la visualisation et de la navigation dans
les donn\'{e}es.

\begin{figure*}[hbt]
  \begin{center}
    \includegraphics*[]{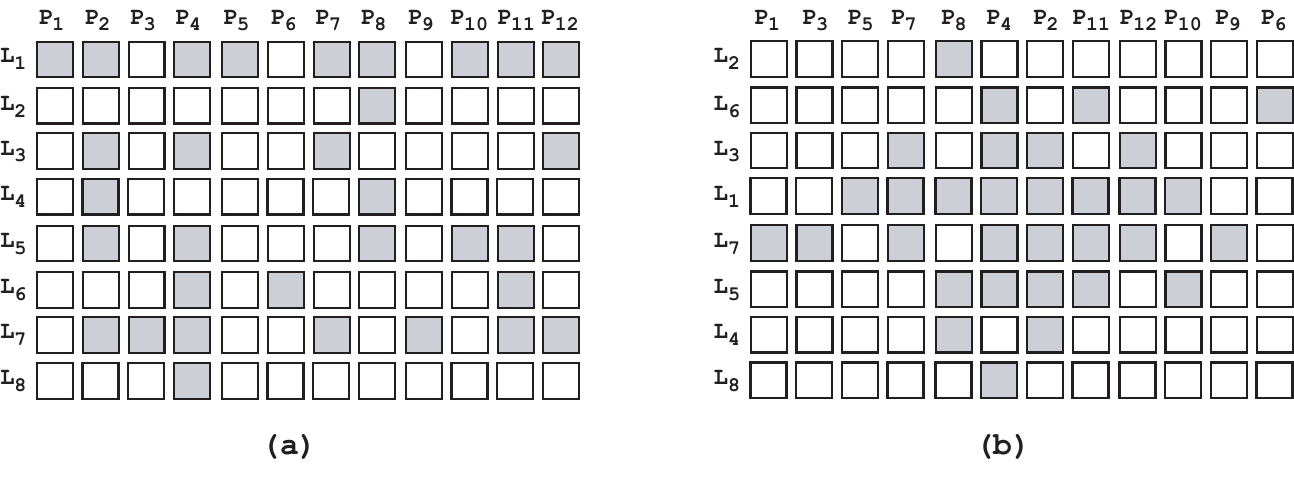}
    \caption{Exemple de deux repr\'{e}sentations d'un espace de donn\'{e}es}
    \label{espaces_representations}
  \end{center}
\end{figure*}

Prenons l'exemple de la figure~\ref{espaces_representations} qui pr\'{e}sente un
cube de donn\'{e}es bancaires \`{a} deux dimensions : les localit\'{e}s g\'{e}ographiques des
agences ($L_{1},\dots,L_{8}$) et les produits de la banque
($P_{1},\dots,P_{12}$). Les cellules gris\'{e}es sur la figure sont pleines et
repr\'{e}sentent la mesure de faits existants (chiffres d'affaires, par exemple)
alors que les cellules blanches sont vides et correspondent \`{a} des faits
inexistants (pas de mesures pour ces croisements de modalit\'{e}s). D'apr\`{e}s la
figure~\ref{espaces_representations}, la r\'{e}partition des cellules pleines dans
la repr\'{e}sentation (a) ne se pr\^{e}te pas facilement \`{a} l'interpr\'{e}tation. En effet,
visuellement, l'information est \'{e}parpill\'{e}e (d'une fa\c{c}on al\'{e}atoire) dans
l'espace de repr\'{e}sentation des donn\'{e}es. En revanche, dans la repr\'{e}sentation
(b), les cellules pleines sont concentr\'{e}es dans la zone centrale du cube. Cette
repr\'{e}sentation offre des possibilit\'{e}s de comparaison et d'analyse des valeurs
des cellules pleines (les mesures des faits) plus ais\'{e}es et plus rapides pour
l'utilisateur.

Notons que les deux repr\'{e}sentations de la figure~\ref{espaces_representations}
correspondent au m\^{e}me cube de donn\'{e}es. La repr\'{e}sentation (b) est obtenue par
simples permutations de lignes et de colonnes de la repr\'{e}sentation (a). Dans la
plupart des serveurs OLAP, les modalit\'{e}s d'une dimension sont pr\'{e}sent\'{e}es selon
un ordre arbitraire. En g\'{e}n\'{e}ral, cet ordre est alphab\'{e}tique pour les libell\'{e}s
des modalit\'{e}s et chronologique pour les dimensions temporelles.
Malheureusement, dans le cas des cubes \'{e}parses et volumineux, ce choix entra\^{\i}ne
des repr\'{e}sentations de donn\'{e}es inadapt\'{e}es \`{a} l'analyse, voire m\^{e}me difficilement
exploitables, comme c'est le cas de la repr\'{e}sentation (a) de la
figure~\ref{espaces_representations}.

La composante visuelle de l'OLAP est primordiale dans un processus d\'{e}cisionnel.
En effet, de la qualit\'{e} et de la clart\'{e} de celle-ci d\'{e}pendent les orientations
de l'utilisateur dans son exploration du cube. Ceci d\'{e}termine la qualit\'{e} des
r\'{e}sultats finaux de l'analyse en ligne. En se basant sur notre id\'{e}e de
l'arrangement des modalit\'{e}s des dimensions illustr\'{e}e dans l'exemple pr\'{e}c\'{e}dent,
nous proposons une m\'{e}thode permettant \`{a} l'utilisateur d'am\'{e}liorer
automatiquement la qualit\'{e} de la repr\'{e}sentation des donn\'{e}es. Nous souhaitons
produire une meilleure visualisation homog\'{e}n\'{e}isant au mieux le nuage des faits
(cellules pleines) et mettant en avant des points de vue int\'{e}ressants pour
l'analyse.

Notre id\'{e}e d'arrangement consiste \`{a} rassembler g\'{e}om\'{e}triquement les cellules
pleines dans l'espace de repr\'{e}sentation des donn\'{e}es. Dans ce travail, nous ne
cherchons pas \`{a} diminuer l'\'{e}parsit\'{e} du cube, mais \`{a} l'organiser de mani\`{e}re
intelligente pour att\'{e}nuer l'impact n\'{e}gatif sur la visualisation qu'elle
engendre. Nous \'{e}valuons l'organisation des donn\'{e}es de notre m\'{e}thode par un
indice de qualit\'{e} de la repr\'{e}sentation des donn\'{e}es que nous d\'{e}finissons dans la
section suivante.

Pour des raisons de complexit\'{e} de traitements, nous avons exclu la recherche
d'un optimum global, voire m\^{e}me local, de l'indice de qualit\'{e} selon une
exploration exhaustive des configurations possibles du cube ; c'est \`{a} dire,
toutes les combinaisons des arrangements possibles des modalit\'{e}s des dimensions
du cube. En effet, consid\'{e}rons le cas d'un cube \`{a} trois dimensions o\`{u} chaque
dimension comporte seulement $10$ modalit\'{e}s. Le nombre de configurations
possibles pour ce cube est \'{e}gal \`{a} ${A}_{10}^{10}\times {A}_{10}^{10} \times
{A}_{10}^{10}=10!\times10! \times 10!\simeq 4,7\cdot10^{19}$.

Afin de parvenir \`{a} un arrangement convenable des modalit\'{e}s du cube, sans passer
par une recherche exhaustive d'un optimum, nous choisissons d'utiliser les
r\'{e}sultats d'une analyse en correspondances multiples (ACM)~\cite{Ben69,LAP00}.
L'ACM est alors consid\'{e}r\'{e}e comme une heuristique appliqu\'{e}e \`{a} la vol\'{e}e aux
donn\'{e}es du cube que l'utilisateur cherche \`{a} visualiser. Les individus et les
variables de l'ACM correspondent respectivement aux faits et aux dimensions du
cube. En construisant des axes factoriels, l'ACM fournit une repr\'{e}sentation
d'associations entre individus et entre variables dans un espace r\'{e}duit. Ces
axes factoriels permettent d'ajuster au mieux le nuage de points des individus
et des variables. Dans le cas de notre approche, afin de mieux repr\'{e}senter les
donn\'{e}es dans un cube, nous proposons d'exploiter les coordonn\'{e}es de ses
modalit\'{e}s sur les axes factoriels. Ces coordonn\'{e}es d\'{e}terminent l'ordre
d'arrangement des modalit\'{e}s dans les dimensions. Cependant, l'ACM s'applique
sur un tableau disjonctif complet obtenu en rempla\c{c}ant dans le tableau initial
chaque variable qualitative par l'ensemble des variables indicatrices des
diff\'{e}rentes modalit\'{e}s de cette variable.

Dans la section suivante, nous formalisons les \'{e}tapes de notre approche. Cette
formalisation pr\'{e}sente la construction du tableau disjonctif complet \`{a} partir
du cube de donn\'{e}es, l'ACM, l'arrangement des modalit\'{e}s des dimensions et
l'indice de qualit\'{e} de la repr\'{e}sentation des donn\'{e}es.

\section{Formalisation}\label{formalisation}

\subsection{Notations}

Dans la suite de l'article, nous consid\'{e}rons $\mathcal{C}$ un cube de donn\'{e}es \`{a}
$d$ dimensions, $m$ mesures et $n$ faits ($d, m, n \in \mathbb{N}^{*}$). Nous
adoptons les notations suivantes :
$D_{1}, \dots , D_{t}, \dots , D_{d}$  repr\'{e}sentent les $d$ dimensions de
$\mathcal{C}$.

Pour la clart\'{e} de l'expos\'{e}, nous supposons que les dimensions ne comportent pas
de hi\'{e}rarchies. Nous consid\'{e}rons que la dimension $D_{t}$ ($t \in \{1, \dots ,
d \}$) est un ensemble de $p_{t}$ modalit\'{e}s qualitatives. On note $a_{j}^{t}$
la $j^{\text{i\`{e}me}}$ modalit\'{e} de la dimension $D_{t}$. Ainsi, l'ensemble des
modalit\'{e}s d'une dimension $D_{t}$ est $\{a_{1}^{t}, \dots, a_{j}^{t}, \dots,
a_{p_{t}}^{t}\}$. Soit $p=\sum_{t=1}^{d} {p_{t}}$ le nombre total de toutes les
modalit\'{e}s des $d$ dimensions du cube $\mathcal{C}$.

Une cellule $A$ dans un cube $\mathcal{C}$ est dite pleine (respectivement,
vide) si elle contient une mesure d'un fait existant (respectivement, ne
contient pas de faits).

\subsection{Aplatissement du cube de donn\'{e}es}

Pour aplatir le cube $\mathcal{C}$, nous le repr\'{e}sentons sous forme
bi-dimensionnelle par un tableau disjonctif complet. Pour chaque dimension
$D_{t}$ ($t \in \{1, \dots, d\}$), nous g\'{e}n\'{e}rons une matrice $Z_{t}$ \`{a} $n$
lignes et $p_{t}$ colonnes. $Z_{t}$ est telle que sa $i^{\text{i\`{e}me}}$ ligne
contenant $(p_{t}-1)$ fois la valeur 0 et une fois la valeur 1 dans la colonne
correspondant \`{a} la modalit\'{e} que prend le fait $i$~($i \in
\{1, \dots, n\}$). 
Le terme g\'{e}n\'{e}ral de la matrice $Z_{t}$ s'\'{e}crit :

$$z^{t}_{ij} = \left\{
\begin{array}{ll}
1 & \text{si le fait $i$ prend la modalit\'{e} $a_{j}^{t}$ de la dimension
$D_{t}$}\\ 0 & \text{sinon}
\end{array}
\right. $$

En juxtaposant les $d$ matrices $Z_{t}$, nous construisons la matrice $Z$ \`{a} $n$
lignes et $p$ colonnes. $Z = [Z_{1}, Z_{2}, \dots, Z_{t}, \dots, Z_{d}]$ est un
tableau disjonctif complet qui d\'{e}crit les $d$ positions des $n$ faits du cube
$\mathcal{C}$ par un codage binaire.

\subsection{Application de l'ACM}

\`A partir du tableau disjonctif complet $Z$, nous construisons le tableau
sym\'{e}trique $B~=~Z'Z$ ($Z'$ d\'{e}signe la transpos\'{e}e de $Z$) d'ordre $(p,p)$, qui
rassemble les croisements deux \`{a} deux de toutes les dimensions du cube
$\mathcal{C}$. $B$ est appel\'{e} tableau de contingence de {\it ``Burt''} associ\'{e}
\`{a} $Z$.

Soit $X$ la matrice diagonale, d'ordre $(p,p)$, ayant les m\^{e}mes \'{e}l\'{e}ments
diagonaux que $B$ et des z\'{e}ros ailleurs. Pour trouver les axes factoriels, nous
diagonalisons la matrice $S~=~\frac{1}{d}Z'ZX^{-1}$ dont le terme g\'{e}n\'{e}ral est :
$$s_{jj'}=\frac{1}{d z_{.j'}}\sum_{i=1}^{n}z_{ij}z_{ij'}$$

Apr\`{e}s diagonalisation, nous obtenons $(p-d)$ valeurs propres de $S$ not\'{e}es
$\lambda_{\alpha}$ ($\alpha\in\{1,\dots,(p-d)\}$). Chaque valeur propre
$\lambda_{\alpha}$ correspond \`{a} un axe factoriel $F_{\alpha}$, de vecteur
directeur $u_{\alpha}$ et v\'{e}rifiant dans $\mathbb{R}^{p}$ l'\'{e}quation :
$$S u_{\alpha} = \lambda_{\alpha} u_{\alpha}$$


Les modalit\'{e}s de la dimension $D_{t}$ sont projet\'{e}es sur les $(p-d)$ axes
factoriels. Soit $\varphi_{\alpha}^{t}$ le vecteur des projections des $p_{t}$
modalit\'{e}s de $D_{t}$ sur $F_{\alpha}$. Notons que
${\varphi_{\alpha}^{t}}^{'}=[\varphi_{\alpha 1}^{t}, \dots, \varphi_{\alpha
j}^{t}, \dots, \varphi_{\alpha p_{t}}^{t}]$.

D\'{e}signons par $\varphi_{\alpha}$ le vecteur des $p$ projections des modalit\'{e}s
de toutes les dimensions sur l'axe factoriel $\alpha$. Notons que
$\varphi_{\alpha}^{'}=[\varphi_{\alpha}^{1}, \dots, \varphi_{\alpha}^{t},
\dots, \varphi_{\alpha}^{p}]$ et que $\varphi_{\alpha}$ v\'{e}rifie l'\'{e}quation :
$$\frac{1}{d}X^{-1}Z'Z\varphi_{\alpha}=\lambda_{\alpha}\varphi_{\alpha}$$


La contribution d'une modalit\'{e} $a_{j}^{t}$ dans la construction de l'axe
$\alpha$ est \'{e}valu\'{e}e par :
$$Cr_{\alpha}(a_{j}^{t})=\frac{z_{.j}^{t} {\varphi_{\alpha
j}^{t}}^{2} }{n d \lambda_{\alpha}}$$

O\`{u} $z_{.j}^{t}=\sum_{i=1}^{n}z_{ij}^{t}$ correspond au nombre de faits dans le
cube $\mathcal{C}$ ayant la modalit\'{e} $a_{j}^{t}$ (poids de la modalit\'{e}
$a_{j}^{t}$ dans le cube).


La contribution d'une dimension $D_{t}$ dans la construction du facteur
$\alpha$ est la somme des contributions des modalit\'{e}s de cette dimension, soit
:
$$Cr_{\alpha}(D_{t})=\sum_{j=1}^{p_{t}}
Cr_{\alpha}(a_{j}^{t}) =\frac{1}{n d \lambda_{\alpha}}
\sum_{j=1}^{p_{t}}z_{.j}^{t} {\varphi_{\alpha j}^{t}}^{2}$$

\subsection{Arrangement des modalit\'{e}s du cube}

Notre id\'{e}e consiste \`{a} associer chaque dimension initiale $D_{t}$ \`{a} un axe
factoriel $F_{\alpha}$. Pour cela, nous exploitons les contributions relatives
des dimensions dans la construction des axes factoriels.

Pour une dimension $D_{t}$ donn\'{e}e, nous cherchons, parmi les axes factoriels
$F_{\alpha}$, celui qui a \'{e}t\'{e} le mieux expliqu\'{e} par les modalit\'{e}s de cette
dimension. Nous cherchons \`{a} maximiser la valeur de
$\lambda_{\alpha}Cr_{\alpha}(D_{t})$. Il s'agit donc de chercher l'axe
$F_{\alpha^{*}}$ pour lequel la somme des carr\'{e}s des projections pond\'{e}r\'{e}es des
modalit\'{e}s de la dimension $D_{t}$ est maximale. Nous cherchons l'indice
$\alpha^{*}$ v\'{e}rifiant l'\'{e}quation suivante :


$$\lambda_{\alpha^{*}}Cr_{\alpha^{*}}(D_{t})=\max_{\alpha\in\{1,\dots,p-d\}}(\lambda_{\alpha}Cr_{\alpha}(D_{t}))$$

\`{A} partir des coordonn\'{e}es des $p_{t}$ projections $\varphi_{\alpha^{*} j}^{t}$
des modalit\'{e}s $a_{j}^{t}$ sur l'axe $F_{\alpha^{*}}$, nous appliquons un tri
croissant de ces coordonn\'{e}es. Ce tri fournit un ordre des indices $j$ selon
lequel nous arrangeons les modalit\'{e}s~$a_{j}^{t}$ de la dimension~$D_{t}$.

L'int\'{e}r\^{e}t de cet arrangement est de converger vers une r\'{e}partition des
modalit\'{e}s de la dimension suivant l'axe factoriel. Cet arrangement a pour effet
de concentrer les cases pleines au centre du cube et d'\'{e}loigner les cases vides
vers les extr\'{e}mit\'{e}s. Sans diminuer l'\'{e}parsit\'{e}, cette m\'{e}thode nous permet
n\'{e}anmoins d'am\'{e}liorer la r\'{e}partition des donn\'{e}es dans le cube. Pour estimer la
qualit\'{e} de cet arrangement, nous proposons un indice pour \'{e}valuer l'homog\'{e}n\'{e}it\'{e}
du cube.

\begin{figure*}[hbt]
  \begin{center}
    \includegraphics*[]{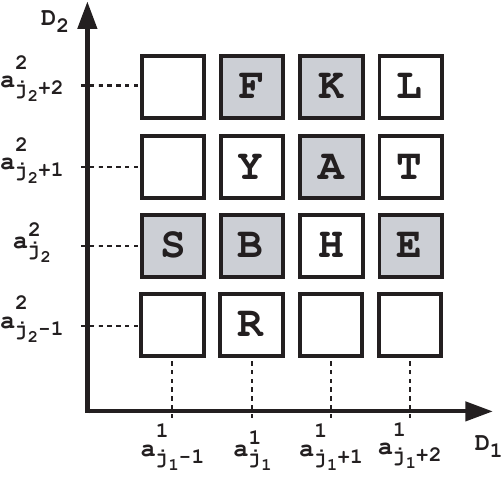}
    \caption{Exemple en 2 dimensions de la notion de voisinage
des cellules d'un cube de donn\'{e}es}
    \label{voisinage}
  \end{center}
\end{figure*}

\subsection{Indice d'homog\'{e}n\'{e}it\'{e}}

Dans cette section, nous proposons un indice permettant de mesurer
l'homog\'{e}n\'{e}it\'{e} de la r\'{e}partition g\'{e}om\'{e}trique des cellules dans un cube. Gr\^{a}ce \`{a}
cet indice, nous pouvons \'{e}valuer le gain induit par l'arrangement des modalit\'{e}s
des dimensions. Nous consid\'{e}rons que plus les cellules pleines (ou bien vides)
sont concentr\'{e}es, plus le cube est dit {\it ``homog\`{e}ne''}.

Une cellule dans un cube repr\'{e}sente une ou plusieurs mesures agr\'{e}g\'{e}es des
faits. Les modalit\'{e}s des dimensions constituent les coordonn\'{e}es des cellules
dans le cube. Soit $A=(a_{j_{1}}^1,\dots,a_{j_{t}}^t,\dots,a_{j_{d}}^d)$ une
cellule dans le cube $\mathcal{C}$, avec $t\in\{1,\dots,d\}$ et
$j_{t}\in\{1,\dots,p_{t}\}$. $j_{t}$ est l'indice de la modalit\'{e} que prend la
cellule $A$ pour la dimension $D_{t}$.

Nous consid\'{e}rons que toutes les modalit\'{e}s des dimensions $D_{t}$ sont
g\'{e}om\'{e}triquement ordonn\'{e}es dans l'espace de repr\'{e}sentation des donn\'{e}es selon
l'ordre des indices $j_{t}$. C'est \`{a} dire, la modalit\'{e} $a_{j_{t}-1}^t$ pr\'{e}c\`{e}de
$a_{j_{t}}^t$, qui, \`{a} son tour, pr\'{e}c\`{e}de $a_{j_{t}+1}^t$ (voir l'exemple de la
figure~\ref{voisinage}). L'ordre des indices $j_{t}$ correspond \`{a} l'ordre dans
lequel sont arrang\'{e}es dans l'espace les modalit\'{e}s de la dimension $D_{t}$. Nous
d\'{e}finissons \`{a} pr\'{e}sent la notion de voisinage pour les cellules d'un cube.

\begin{defn}[Cellules voisines]
Soit $A=(a_{j_{1}}^1,\dots,a_{j_{t}}^t,\dots,a_{j_{d}}^d)$ une cellule dans un
cube $\mathcal{C}$. La cellule
$B=(b_{j_{1}}^1,\dots,b_{j_{t}}^t,\dots,b_{j_{d}}^d)$ est dite voisine de $A$,
not\'{e}e $B \dashv A$, si $\forall t\in\{1,\dots,d\}$, les coordonn\'{e}es de $B$
v\'{e}rifient : $b_{j_{t}}^t~=~a_{j_{t}-1}^t$ ou $b_{j_{t}}^t~=~a_{j_{t}}^t \text{
ou } b_{j_{t}}^t=a_{j_{t}+1}^t$. Exception faite du cas o\`{u} $\forall
t\in\{1,\dots,d\}\ b_{j_{t}}^t=a_{j_{t}}^t$, $B$ n'est pas consid\'{e}r\'{e}e comme une
cellule voisine de $A$ car $B=A$.
\end{defn}

Dans l'exemple de la figure~\ref{voisinage}, la cellule $B$ est voisine de $A$
($B \dashv A$). $Y$ est aussi voisine de $A$ ($Y \dashv A$). En revanche, les
cellules $S$ et $R$ ne sont pas voisines de $A$. Ceci nous ram\`{e}ne \`{a} d\'{e}finir le
voisinage d'une cellule.

\begin{defn}[Voisinage d'une cellule]
Soit $A$ une cellule du cube $\mathcal{C}$, nous d\'{e}finissons le voisinage de
$A$, not\'{e} $\mathcal{V}(A)$, par l'ensemble de toutes les cellules $B$ de
$\mathcal{C}$ qui sont voisines de $A$.
$$\mathcal{V}(A)=\{B\in\mathcal{C} \text{ tel que } B \dashv A\}$$
\end{defn}

Par exemple, dans la figure~\ref{voisinage}, le voisinage de la cellule $A$
correspond \`{a} l'ensemble $\mathcal{V}(A)=\{F, K, L, Y, T, B, H, E\}$.

\begin{defn}[Fonction $\Delta$]
Nous d\'{e}finissons une fonction $\Delta$ de $\mathcal{C}$ dans $\mathbb{N}$ tel
que :
$$\forall A\in\mathcal{C},\  \Delta(A)=\sum_{B\in\mathcal{V}(A)}\delta(A,B)$$

Avec $\delta$ est une fonction d\'{e}finie comme suit :

$$
\begin{array}{cll}
\delta : \mathcal{C}\times\mathcal{C} & \longrightarrow & \mathbb{N}
\\
\delta(A,B) & \longmapsto & \left \{
\begin{array}{ll}
1 & \text{si $A$ et $B$ sont pleines}\\ 0 & \text{sinon}
\end{array}
\right.
\end{array}
$$

$\Delta(A)$ correspond au nombre de cellules pleines et voisines de $A$.
\end{defn}

En supposant que les cellules grises repr\'{e}sentent les cellules pleines dans la
figure~\ref{voisinage}, $\Delta(A)=4$ puisque $F$, $K$, $B$ et $E$ sont les
seules cellules qui sont \`{a} la fois pleines et voisines de $A$.

\begin{defn}[Indice d'homog\'{e}n\'{e}it\'{e} brut]
Nous d\'{e}finissons l'indice d'homog\'{e}n\'{e}it\'{e} brut d'un cube $\mathcal{C}$, not\'{e}
$IHB(\mathcal{C})$, par la somme de tous les couples de ses cellules qui sont \`{a}
la fois pleines et voisines.
$$IHB(\mathcal{C})=\sum_{A\in\mathcal{C}}\Delta(A)$$
\end{defn}

Par exemple, l'indice d'homog\'{e}n\'{e}it\'{e} brut du cube de la figure~\ref{voisinage}
se calcule comme suit :
$$IHB(\mathcal{C})=\Delta(F)+\Delta(K)+\Delta(A)+\Delta(S)+\Delta(B)+\Delta(E)=2+2+4+1+2+1=12$$

La meilleure repr\'{e}sentation d'un cube de donn\'{e}es correspond au cas o\`{u} ce
dernier est compl\`{e}tement non vide. C'est \`{a} dire, toutes ses cellules sont
pleines. Dans ce cas, l'indice d'homog\'{e}n\'{e}it\'{e} brut est maximal :
$$IHB_{max}(\mathcal{C})=\sum_{A\in\mathcal{C}}\sum_{B\in\mathcal{V}(A)}1$$

\begin{defn}[Indice d'homog\'{e}n\'{e}it\'{e}]
Nous d\'{e}finissons, l'indice d'homog\'{e}n\'{e}it\'{e} d'un cube $\mathcal{C}$, not\'{e}
$IH(\mathcal{C})$, par le rapport de l'indice de l'homog\'{e}n\'{e}it\'{e} brut sur celui
de l'homog\'{e}n\'{e}it\'{e} maximale.
$$IH(\mathcal{C})=\frac{IHB(\mathcal{C})}{IHB_{max}(\mathcal{C})}=
\frac{\displaystyle{\sum_{A\in\mathcal{C}}}\Delta(A)}
{\displaystyle{\sum_{A\in\mathcal{C}}\sum_{B\in\mathcal{V}(A)}1}}$$
\end{defn}

Apr\`{e}s calcul, l'homog\'{e}n\'{e}it\'{e} maximale du cube exemple de la
figure~\ref{voisinage} \'{e}tant \'{e}gale \`{a} 48, l'indice d'homog\'{e}n\'{e}it\'{e} de ce dernier
est donc $IH(\mathcal{C})=\frac{12}{48}\simeq14,28\%$

Pour mesurer l'apport de l'arrangement des modalit\'{e}s sur la repr\'{e}sentation du
cube de donn\'{e}es, nous calculons le gain en homog\'{e}n\'{e}it\'{e} not\'{e} $g$ selon la
formule :

$$g=\frac{IH(\mathcal{C_{\textit{arr}}})-IH(\mathcal{C_{\textit{ini}}})}{IH(\mathcal{C_{\textit{ini}}})}$$

o\`{u} $IH(\mathcal{C_{\textit{ini}}})$ est l'indice d'homog\'{e}n\'{e}it\'{e} de la
repr\'{e}sentation du cube initial et $IH(\mathcal{C_{\textit{arr}}})$ est celui de
la repr\'{e}sentation arrang\'{e}e selon notre m\'{e}thode. Notons que quelle que soit la
repr\'{e}sentation initiale du cube, l'arrangement fourni en sortie par notre
m\'{e}thode est identique puisque l'ACM n'est pas sensible \`{a} l'ordre des variables
donn\'{e}es en entr\'{e}e.

%

\section{\'Etude de cas}\label{etude}

Pour tester et valider l'approche que nous proposons, nous utilisons un jeu de
donn\'{e}es bancaires extrait du syst\`{e}me d'information du {\it Cr\'{e}dit Lyonnais}. \`{A}
partir de ces donn\'{e}es, nous avons construit un contexte d'analyse (cube de
donn\'{e}es). Un fait du cube correspond au comportement d'achat d'un client. Nous
disposons dans ce cube de $n = 311\:959$ comportements de clients mesur\'{e}s par
le produit net bancaire ($M_{1}$) et le montant des avoirs ($M_{2}$). Le
tableau~\ref{tab:descritption} d\'{e}taille la description des dimensions
consid\'{e}r\'{e}es pour observer ces mesures.

\begin{table}[t]
{\small
\begin{center}
\begin{tabular}{|p{2cm}|p{2cm}|p{8cm}|}
\hline
\textbf{Dimension} & \textbf{Nombre de modalit\'{e}s}  & \textbf{Description} \\
\hline $D_{1}$ : cat\'{e}gorie socio-professionnelle & $p_{1}=58$ & profil
professionnel du client \\ \hline $D_{2}$ : produit & $p_{2}=25$ & d\'{e}tention de
formule(s) qui sont des offres combin\'{e}es de produits
bancaires \\
\hline $D_{3}$ : unit\'{e} commerciale & $p_{3}=65$ & localisations g\'{e}ographiques de vente \\
\hline $D_{4}$ : segment & $p_{4}=15$ & potentiel commercial du client \\
\hline $D_{5}$ : \^{a}ge & $p_{5}=12$ & variable discr\'{e}tis\'{e}e selon
  des tranches d'\^{a}ge de dix ans ([0-10], [11-20], [21-30], etc.) \\
  \hline $D_{6}$ : situation familiale & $p_{6}=6$ & exemple : mari\'{e}, divorc\'{e}, etc. \\
  \hline $D_{7}$ : type client & $p_{7}=4$ & origine du
  client (par exemple, client membre du personnel du Cr\'{e}dit Lyonnais) \\
  \hline $D_{8}$ : march\'{e} & $p_{8}=4$ & une vente r\'{e}alis\'{e}e aupr\`{e}s d'un client est faite sur le march\'{e} ``particulier des professionnels '' si le client est artisan ou exerce une profession lib\'{e}rale,
  etc., ou sur le march\'{e} ``particulier'' sinon \\ \hline
\end{tabular}
\end{center}
} \caption{Description des dimensions du cube exemple}\label{tab:descritption}
\end{table}



\begin{figure}[h]
{\centering \resizebox*{0.78\textwidth}{!}{\includegraphics{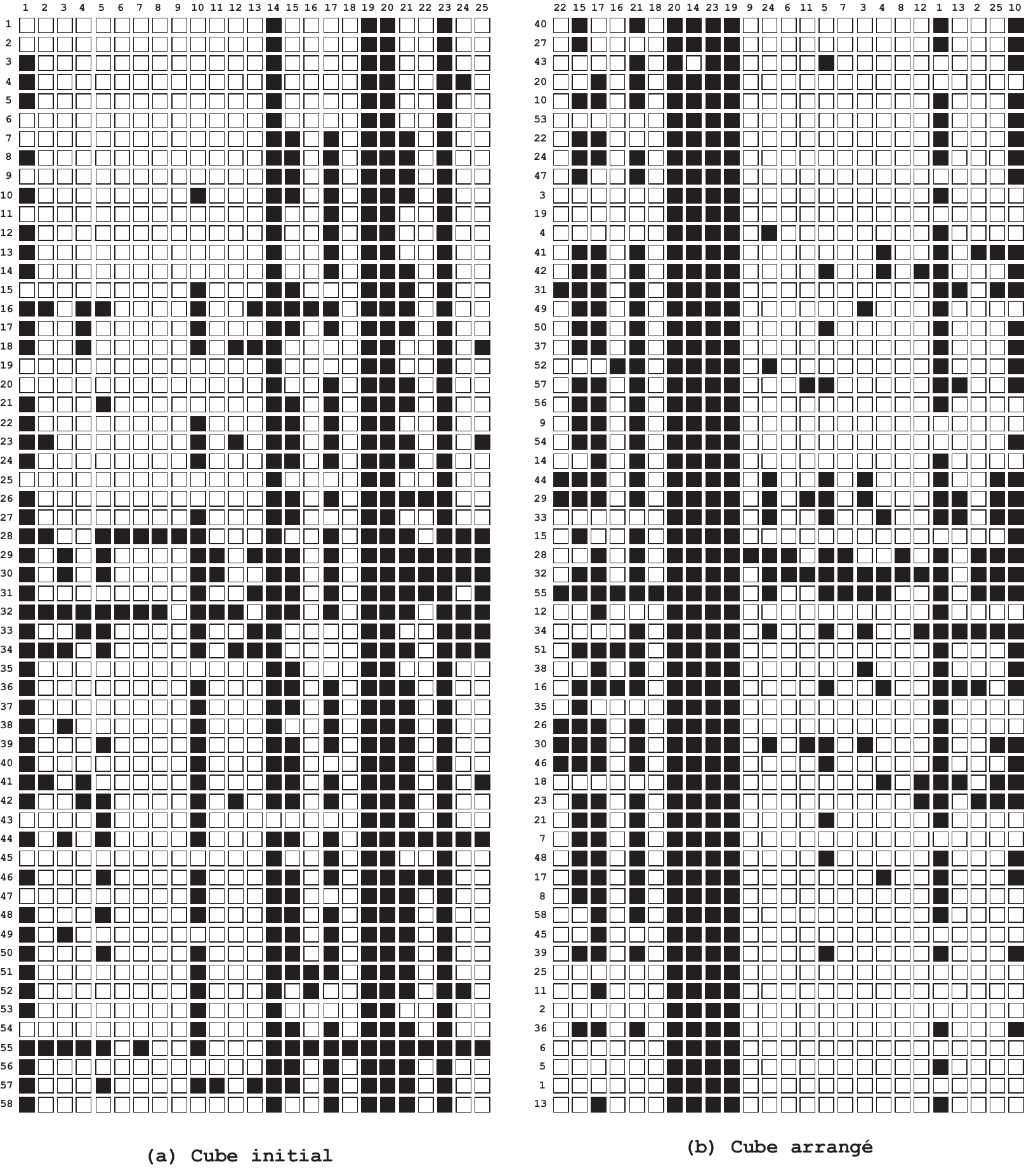}}
\par}
\caption{Le cube de donn\'{e}es avant et apr\`{e}s arrangement des
modalit\'{e}s}\label{exemple_arrang}
\end{figure}

Pour rendre plus claire la suite de notre expos\'{e}, notre \'{e}tude de cas porte sur
un cube \`{a} deux dimensions ($d=2$) : la dimension ``cat\'{e}gorie
socio-professionnelle'' ($D_{1}$) et la dimension ``produit'' ($D_{2})$. La
mesure observ\'{e}e est ``le montant des avoirs''. Nous g\'{e}n\'{e}rons les matrices
$Z_{1}$ et $Z_{2}$ selon un codage binaire disjonctif des modalit\'{e}s des deux
dimensions. Le tableau disjonctif complet $Z=[Z_{1},Z_{2}]$ a $n = 311\:959$
lignes et $p=p_{1}+p_{2}=83$ colonnes.

En appliquant l'ACM sur le tableau $Z$, on obtient $p-d=81$ axes factoriels
$F_{\alpha}$. Chaque axe est caract\'{e}ris\'{e} par sa valeur propre
$\lambda_{\alpha}$ et les contributions apport\'{e}es par les dimensions :
$Cr_{\alpha}(D_{1})$ et $Cr_{\alpha}(D_{2})$. Nous cherchons, pour chaque
dimension, l'axe qui est le mieux contribu\'{e} par cette derni\`{e}re. Nous obtenons
les r\'{e}sultats suivants~:

\begin{itemize}
\item Pour la dimension $D_{1}$,
$\lambda_{45}Cr_{45}(D_{1})=\max_{\alpha\in\{1,\dots,81\}}(\lambda_{\alpha}Cr_{\alpha}(D_{1}))$,
avec $\lambda_{45}=0.5$ et $Cr_{45}(D_{1})=99.9\%$
\item Pour la
dimension $D_{2}$,
$\lambda_{1}Cr_{1}(D_{2})=\max_{\alpha\in\{1,\dots,81\}}(\lambda_{\alpha}Cr_{\alpha}(D_{2}))$,
avec $\lambda_{1}=0.83$ et $Cr_{1}(D_{2})=50\%$.
\end{itemize}

Ainsi, la dimension $D_{1}$ est associ\'{e}e \`{a} l'axe $F_{45}$ et $D_{2}$ \`{a} l'axe
$F_{1}$. Les modalit\'{e}s de $D_{1}$ (respectivement, $D_{2}$) sont arrang\'{e}es
suivant l'ordre croissant de leur projections sur $F_{45}$ (respectivement,
$F_{1}$). Dans la figure~\ref{exemple_arrang}, nous pr\'{e}sentons le r\'{e}sultat de
cet arrangement. La repr\'{e}sentation (a) correspond \`{a} l'arrangement initial du
cube selon l'ordre alphab\'{e}tique des libell\'{e}s des modalit\'{e}s. La repr\'{e}sentation
(b) correspond \`{a} l'arrangement obtenu par l'ordre croissant des projections des
modalit\'{e}s sur les axes factoriels suscit\'{e}s. Pour des raisons de
confidentialit\'{e}, nous masquons les libell\'{e}s des modalit\'{e}s de chaque dimension
ainsi que les valeurs des mesures. Nous rempla\c{c}ons les libell\'{e}s par des codes
chiffr\'{e}s et les mesures existantes par des cases noires. Les cases blanches du
cube repr\'{e}sentent les creux correspondant \`{a} des croisements vides. Sur cet
exemple, le taux d'\'{e}parsit\'{e} du cube~\footnote{Le taux d'\'{e}parsit\'{e} est \'{e}gal au
rapport entre le nombre de cases vides et le nombre total des cases du cube.}
est \'{e}gal \`{a} $64\%$. La valeur de l'indice d'homog\'{e}n\'{e}it\'{e} est de $17,75\%$ pour la
repr\'{e}sentation (a) et de $20,60\%$ pour la repr\'{e}sentation (b). Nous obtenons
donc un gain en homog\'{e}n\'{e}it\'{e} de $16,38\%$ par rapport \`{a} la repr\'{e}sentation
initiale du cube.


Nous avons \'{e}galement appliqu\'{e} notre m\'{e}thode sur un cube \`{a} trois dimensions :
``cat\'{e}gorie socio-professionnelle'' ($D_{1}$), ``produit'' ($D_{2})$ et ``\^{a}ge''
($D_{5}$). Ce cube, dont le taux d'\'{e}parsit\'{e} est \'{e}gal \`{a} $87,94\%$, contient plus
de cellules vides compar\'{e} au cube pr\'{e}c\'{e}dent. L'arrangement des modalit\'{e}s
correspond \`{a} l'ordre alphab\'{e}tique pour $D_{1}$ et $D_{2}$, et \`{a} l'ordre
croissant des tranches d'\^{a}ge pour $D_{5}$. Le cube initial a un indice
d'homog\'{e}n\'{e}it\'{e} de $5,12\%$. Le cube arrang\'{e}, selon notre m\'{e}thode, a un indice
d'homog\'{e}n\'{e}it\'{e} de $6,11\%$. Nous obtenons ainsi un gain de $19,33\%$.

\begin{figure}[h]
{\centering \resizebox*{1\textwidth}{!}{\includegraphics{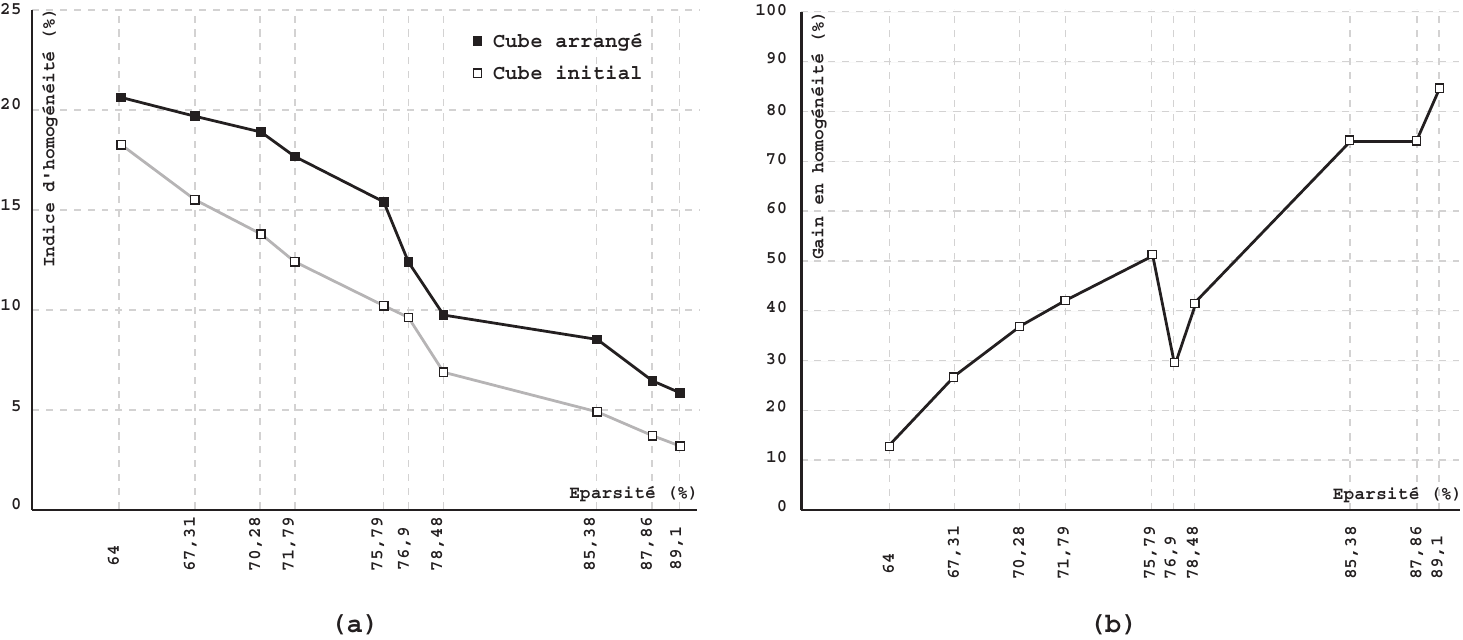}}
\par}
\caption{\'{E}volutions de l'indice d'homog\'{e}n\'{e}it\'{e} et du gain en fonction de
l'\'{e}parsit\'{e}}\label{experimentation}
\end{figure}

Nous avons r\'{e}alis\'{e} une s\'{e}rie d'exp\'{e}rimentations de notre m\'{e}thode sur le premier
cube (le cube \`{a} deux dimension $D_{1}$ et $D_{2}$), pour diff\'{e}rentes valeurs du
taux d'\'{e}parsit\'{e}. Afin de mesurer l'impact de l'\'{e}parsit\'{e} sur notre m\'{e}thode, nous
avons tir\'{e} plusieurs \'{e}chantillons al\'{e}atoires \`{a} partir de la population du cube
initial (les $n$ faits du cube). En variant le taux d'\'{e}chantillonnage, nous
parvenons \`{a} faire varier l'\'{e}parsit\'{e} du cube.

La figure~\ref{experimentation}~(a) montre l'\'{e}volution de l'indice
d'homog\'{e}n\'{e}it\'{e} du cube initial et du cube arrang\'{e} en fonction de l'\'{e}parsit\'{e}.
Nous remarquons que les valeurs de l'indice sont d\'{e}croissantes en fonction de
l'\'{e}parsit\'{e} du cube. Ceci est naturellement d\^{u} \`{a} la construction de cet indice
qui d\'{e}pend fortement du nombre de cellules pleines dans le cube. Notons aussi
que, quelle que soit l'\'{e}parsit\'{e}, le cube obtenu par arrangement selon notre
m\'{e}thode est toujours de meilleure qualit\'{e} que le cube initial au sens de notre
indice d'homog\'{e}n\'{e}it\'{e}. Dans tous les cas, nous r\'{e}alisons un gain en homog\'{e}n\'{e}it\'{e}
lors de l'arrangement du cube.

D'apr\`{e}s la figure~\ref{experimentation}~(b), le gain en homog\'{e}n\'{e}it\'{e} a une
tendance g\'{e}n\'{e}rale croissante en fonction de l'\'{e}parsit\'{e} du cube. En effet, plus
le cube est \'{e}parse, plus nous avons une meilleure marge de man\oe uvre pour
concentrer les donn\'{e}es et les regrouper ensemble autour des axes factoriels de
l'ACM.

Notons aussi que le gain en homog\'{e}n\'{e}it\'{e}, qui est toujours positif, peut fl\'{e}chir
localement (voir figure~\ref{experimentation}~(b)). Ceci est inh\'{e}rent \`{a} la
structure des donn\'{e}es. C'est \`{a} dire, si les donn\'{e}es du cube initial sont d\'{e}j\`{a}
dans une repr\'{e}sentation homog\`{e}ne, l'application de notre m\'{e}thode n'apportera
pas de gain consid\'{e}rable. En effet, dans ce cas, la m\'{e}thode n'aura qu'un effet
de translation du nuage des fait vers les zones centrales des axes factoriels.

\section{Travaux connexes}\label{discussion}

L'am\'{e}lioration de l'espace de repr\'{e}sentation des donn\'{e}es multidimensionnelles
dans l'OLAP a fait l'objet de plusieurs travaux de recherche. Rappelons que,
dans notre cas, cette am\'{e}lioration se traduit par la concentration des donn\'{e}es
autour des axes factoriels d'une ACM. Cela a pour effet de produire une
meilleure visualisation homog\'{e}n\'{e}isant au mieux le nuage des faits et mettant en
avant des points de vue int\'{e}ressants pour l'analyse.

Les travaux de recherche qui se sont int\'{e}ress\'{e}s \`{a} l'\'{e}tude de l'espace de
repr\'{e}sentation ont \'{e}t\'{e} men\'{e}s suite \`{a} des motivations diff\'{e}rentes. Tandis que
certains se sont pench\'{e}s sur des aspects d'optimisation technique (stockage,
temps de r\'{e}ponse, etc.), d'autres s'int\'{e}ressent plut\^{o}t \`{a} l'aspect de l'analyse
en ligne, et particuli\`{e}rement \`{a} la visualisation. Notre travail s'articule
davantage autour des seconds travaux. Tout d'abord, nous pr\'{e}sentons les travaux
ayant trait\'{e} l'approximation des cubes de donn\'{e}es, leur compression et
l'optimisation des calculs d'agr\'{e}gats.




En se basant sur le principe d'approximation par ondelettes
(\textit{wavelets}), Vitter {\it et al.}~\cite{ViWa99} proposent un algorithme
pour construire un cube de donn\'{e}es compact. L'algorithme propos\'{e} fournit des
r\'{e}sultats meilleurs que ceux de l'approximation par histogrammes ou par
\'{e}chantillonnage al\'{e}atoire~\cite{VWI98}. Dans le m\^{e}me ordre d'id\'{e}es, Barbara et
Sullivan~\cite{BaSu97} ont propos\'{e} l'approche \textbf{\texttt{Quasi-Cube}} qui,
au lieu de mat\'{e}rialiser la totalit\'{e} d'un cube, mat\'{e}rialise une partie de ce
dernier en se basant sur une description incompl\`{e}te mais suffisante de ses
donn\'{e}es. Les donn\'{e}es non mat\'{e}rialis\'{e}es sont ensuite approxim\'{e}es par une
r\'{e}gression lin\'{e}aire.


Une technique de compression bas\'{e}e sur la mod\'{e}lisation statistique de la
structure des donn\'{e}es d'un cube a \'{e}t\'{e} propos\'{e}e dans~\cite{SFB99}. Apr\`{e}s
estimation de la densit\'{e} de probabilit\'{e} des donn\'{e}es, les auteurs construisent
une repr\'{e}sentation compacte des donn\'{e}es capable de supporter des requ\^{e}tes
d'agr\'{e}gation. Cette technique n'a de sens que dans le cas de cubes pr\'{e}sentant
des dimensions continues.

La m\'{e}thode de compression \textbf{\texttt{Dwarf}} propos\'{e}e dans~\cite{SDR02},
r\'{e}duit l'espace de stockage d'un cube de donn\'{e}es. Cette m\'{e}thode consiste \`{a}
identifier les n-uplets redondants dans la table de faits. Les redondances de
donn\'{e}es sont ensuite remplac\'{e}es par un seul enregistrement. Wang {\it et
al.}~\cite{WFL02} proposent de factoriser ces redondances par un seul n-uplet
de base appel\'{e} \textbf{\texttt{BST}} ({\it Base Single Tuple}). \`A partir du
\textbf{\texttt{BST}}, les auteurs construisent un cube de donn\'{e}es de moindre
taille \textbf{\texttt{MinCube}} (\textit{Minimal condensed BST Cube}). Cette
approche requiert des temps de traitement relativement longs. En vue de
rem\'{e}dier \`{a} cette limite, Feng {\it et al.}~\cite{FFD04} ont repris l'approche
en introduisant une nouvelle structure de donn\'{e}es \textbf{\texttt{PrefixCube}}.
Ils sugg\`{e}rent de ne plus utiliser tous les \texttt{\textbf{BST}} dans la
construction du cube mais plut\^{o}t de se contenter d'un seul
\textbf{\texttt{BST}} par dimension. En contre partie, ils proposent
l'algorithme \textbf{\texttt{BU-BST}} pour la construction d'un cube compress\'{e}
(\textit{Bottom Up BST algorithm}). Cet algorithme est une version am\'{e}lior\'{e}e de
l'algorithme \textbf{\texttt{BUC}} (\textit{Bottom Up Computation algorithm})
propos\'{e} \`{a} l'origine dans~\cite{BeRa99}.

Lakshmanan \textit{et al.}~\cite{LPH02} proposent la m\'{e}thode
\textbf{\texttt{Quotient Cube}} pour la compression d'un cube de donn\'{e}es en
r\'{e}sumant son contenu s\'{e}mantique et en le structurant sous forme de partitions
de classes. La meilleure partition n'est pas seulement celle qui permet de
r\'{e}duire la taille du cube mais aussi celle qui permet de conserver une
structure de treillis valide donnant la possibilit\'{e} de naviguer avec les
op\'{e}rations d'agr\'{e}gation (\textit{Roll-Up}) et de sp\'{e}cification
(\textit{Drill-Down}) dans le cube r\'{e}duit. Malheureusement, la technique des
\textbf{\texttt{Quotient Cube}} fournit des structures peu compactes. De plus,
ces structures ne sont pas adapt\'{e}es aux mises \`{a} jours des donn\'{e}es.
Dans~\cite{LPZ03}, Lakshmanan et \textit{al.} proposent une nouvelle version
am\'{e}lior\'{e}e \textbf{\texttt{QC-Tree}} (\textit{Quotient Cube Tree}) qui pallie
les limites de la technique des \textbf{\texttt{Quotient Cube}}.
\textbf{\texttt{QC-Tree}} permet de rechercher les structures compactes de
donn\'{e}es dans un cube, d'extraire et de construire les cubes int\'{e}ressants \`{a}
partir des donn\'{e}es mises \`{a} jour.

Feng \textit{et al.}~\cite{FAE03} proposent la m\'{e}thode \textbf{\texttt{Range
CUBE}} pour la compression des cubes en se basant sur les corr\'{e}lations entre
les cellules du cube. Cette approche consiste \`{a} cr\'{e}er un arrangement des
cellules d'un cube selon un certain formalisme d'appartenance introduit dans
les n\oe uds du treillis du cube original. Cet arrangement permet de produire
une nouvelle structure du cube plus compacte et moins co\^{u}teuse en stockage et
en temps de r\'{e}ponse.


Ross et Srivastava~\cite{RoSr97} traitent le probl\`{e}me de l'optimisation du
calcul d'agr\'{e}gats dans les cubes de donn\'{e}es \'{e}parses. Les auteurs proposent
l'algorithme \textbf{\texttt{Partitioned-Cube}} qui partitionnent les relations
entre les donn\'{e}es d'un cube en plusieurs fragments de fa\c{c}on \`{a} ce qu'ils
tiennent en m\'{e}moire centrale. Cette mesure permet de r\'{e}duire le co\^{u}t des
entr\'{e}es/sorties. Les fragments de donn\'{e}es sont ensuite trait\'{e}s ind\'{e}pendamment,
un par un, afin de calculer les agr\'{e}gats possibles et de g\'{e}n\'{e}rer des sous-cubes
de donn\'{e}es. Cette notion de fragment est reprise dans les travaux de Li
\textit{et al.}~\cite{LHG04}. Leur m\'{e}thode, appel\'{e}e \textbf{\texttt{Shell
Fragment}}, partitionne un ensemble de donn\'{e}es de forte dimensionnalit\'{e} en
sous-ensembles disjoints de donn\'{e}es de dimensionnalit\'{e}s moins importantes
appel\'{e}s {\it ``fragments''}. Pour chaque fragment est calcul\'{e} un cube de
donn\'{e}es local. Les identifiants des n-uplets participant \`{a} la construction de
cellules non vides dans un fragment sont enregistr\'{e}s. Ces identifiants sont
utilis\'{e}s pour lier diff\'{e}rents fragments et reconstruire de petits cubes
(cubo\"{\i}des) n\'{e}cessaires \`{a} l'\'{e}valuation d'une requ\^{e}te. Le cube de donn\'{e}es de
d\'{e}part est assembl\'{e} via ces fragments.

Enfin, citons les travaux de Choong {\it et al.} \cite{CLL04,CLM03} qui ont une
motivation similaire \`{a} la n\^{o}tre. Les auteurs utilisent les r\`{e}gles floues
(combinaison d'un algorithme de r\`{e}gles d'association et de la th\'{e}orie des
sous-ensembles flous) afin de faciliter la visualisation et la navigation dans
l'espace de repr\'{e}sentation des cubes de donn\'{e}es. Leur approche, consiste \`{a}
identifier et \`{a} construire des blocs de donn\'{e}es similaires au sens de la mesure
du cube. Cependant, cette approche ne prend pas en compte le probl\`{e}me
d'\'{e}parsit\'{e} du cube. De plus, elle se base sur le comptage du nombre
d'occurrences des mesures o\`{u} ces derni\`{e}res sont consid\'{e}r\'{e}es comme des nombres
entiers.

\section{Conclusion et perspectives}\label{conclusion}
Dans cet article, nous avons propos\'{e} une nouvelle approche apportant une
solution au probl\`{e}me de la visualisation des donn\'{e}es dans un cube \'{e}parse. Sans
r\'{e}duire l'\'{e}parsit\'{e}, nous cherchons \`{a} organiser l'espace multidimensionnel des
donn\'{e}es afin de regrouper g\'{e}om\'{e}triquement les cellules pleines dans un cube. La
recherche d'un arrangement optimal du cube est un probl\`{e}me complexe et co\^{u}teux
en temps de calcul. Nous avons choisi d'utiliser les r\'{e}sultats de l'ACM comme
heuristique pour r\'{e}duire cette complexit\'{e}. Notre approche consiste \`{a} arranger
les modalit\'{e}s des dimensions d'un cube, selon les besoins d'analyse de
l'utilisateur, en fonction des r\'{e}sultats fournis par l'ACM. Pour \'{e}valuer
l'apport de cette nouvelle repr\'{e}sentation de donn\'{e}es, nous avons propos\'{e} un
indice d'homog\'{e}n\'{e}it\'{e} bas\'{e} sur le voisinage. La comparaison des valeurs de
l'indice entre les repr\'{e}sentations initiale et arrang\'{e}e du cube nous permet
d'\'{e}valuer l'efficacit\'{e} de notre approche. Les diff\'{e}rents tests sur notre jeu de
donn\'{e}es bancaires nous ont montr\'{e}, que quelle que soit l'\'{e}parsit\'{e}, notre
approche est pertinente. Le gain en homog\'{e}n\'{e}it\'{e} est croissant en fonction de
l'\'{e}parsit\'{e} et son amplitude est \'{e}galement inh\'{e}rente \`{a} la structure des donn\'{e}es.

Suite \`{a} ce travail, plusieurs perspectives sont \`{a} pr\'{e}voir. Tout d'abord, nous
devons \'{e}tudier la complexit\'{e} de notre m\'{e}thode. Cette \'{e}tude doit prendre en
compte aussi bien les propri\'{e}t\'{e}s du cube (taille, \'{e}parsit\'{e}, cardinalit\'{e}s, etc.)
que l'impact de l'\'{e}volution des donn\'{e}es (rafra\^{\i}chissement de l'entrep\^{o}t de
donn\'{e}es).

Ensuite, \`{a} ce stade de nos travaux, pour appliquer l'ACM, nous tenons seulement
compte de la pr\'{e}sence/absence des faits du cube dans la construction des axes
factoriels. Nous envisageons alors d'introduire les valeurs des mesures comme
pond\'{e}rations des faits (poids des individus de l'ACM). Ceci permettra de
construire des axes factoriels qui traduisent mieux la repr\'{e}sentation des faits
du cube selon leur ordre de grandeur. Dans ce cas, il serait \'{e}galement
int\'{e}ressant d'introduire la notion de distance entre cellules voisines en
fonction des valeurs des mesures qu'elles contiennent.

Dans le m\^{e}me ordre d'id\'{e}es de la pr\'{e}sente m\'{e}thode, nous souhaitons utiliser les
r\'{e}sultats de l'ACM afin de faire \'{e}merger des r\'{e}gions int\'{e}ressantes \`{a} l'analyse
\`{a} partir d'un cube de donn\'{e}es initial. En effet, l'ACM permet de concentrer
dans les zones centrales des axes factoriels les individus ayant un
comportement normal, et d'\'{e}loigner ceux ayant des comportements atypiques vers
les zones extr\^{e}mes. Nous pouvons d\'{e}j\`{a} exploiter les r\'{e}sultats de l'arrangement
des modalit\'{e}s du cube dans le cadre de la distinction de r\'{e}gions correspondant
\`{a} ces comportements caract\'{e}ristiques.

Nous voulons aussi comparer la visualisation obtenue par notre approche avec
celle propos\'{e}e dans~\cite{ChRi98}. Cette derni\`{e}re repr\'{e}sente les r\'{e}sultats
d'une analyse factorielle sous forme d'un diagramme de Bertin~\cite{Ber81} qui
est plus facile \`{a} interpr\'{e}ter. L'objectif de cette m\'{e}thode est de proposer une
visualisation optimis\'{e}e d'un tableau de contingence. Cependant, elle se limite
\`{a} des tableaux \`{a} deux dimensions sans donn\'{e}es manquantes et ne peut pas
s'appliquer \`{a} des cubes \`{a} forte dimensionnalit\'{e}. Notre approche peut \^{e}tre
consid\'{e}r\'{e}e comme une extension de cette m\'{e}thode concernant la dimensionnalit\'{e}
du cube et de l'\'{e}parsit\'{e} de ses donn\'{e}es.

Par ailleurs, la mat\'{e}rialisation des cubes de donn\'{e}es permet le pr\'{e}-calcul et
le stockage des agr\'{e}gats multidimentionnels de mani\`{e}re \`{a} rendre l'analyse OLAP
performante. Cela requiert un temps de calcul important et g\'{e}n\`{e}re un volume de
donn\'{e}es \'{e}lev\'{e} lorsque le cube mat\'{e}rialis\'{e} est \`{a} forte dimentionnalit\'{e}. Au lieu
de calculer la totalit\'{e} du cube, il serait judicieux de calculer et de
mat\'{e}rialiser que les parties int\'{e}ressantes du cube (fragments contenant
l'information utile).  Comme l'information r\'{e}side dans les cellules pleines, le
cube arrang\'{e} obtenu par l'application de l'ACM serait un point de d\'{e}part pour
d\'{e}terminer ces fragments. Ainsi, comme dans~\cite{BaSu97}, chaque fragment
donnera lieu \`{a} un cube local. Les liens entre ces cubes permettront de
reconstruire le cube initial.

Enfin, dans ce travail, nous avons d\'{e}lib\'{e}r\'{e}ment omis de pr\'{e}ciser l'origine de
ces donn\'{e}es. Classiquement, ces donn\'{e}es peuvent \^{e}tre issues d'un entrep\^{o}t de
donn\'{e}es. Mais nous envisageons d'appliquer cette approche dans un contexte
d'entreposage virtuel. Nous entendons par entreposage virtuel la construction
de cube \`{a} la vol\'{e}e \`{a} partir de donn\'{e}es fournies par un syst\`{e}me de m\'{e}diation. Un
enjeu prometteur de notre m\'{e}thode est donc de pouvoir soumettre \`{a}
l'utilisateur, dans le contexte de l'entreposage virtuel, des repr\'{e}sentations
visuellement int\'{e}ressantes des cubes de donn\'{e}es. Selon cette d\'{e}marche,
l'utilisateur est de plus en plus impliqu\'{e} dans le processus d\'{e}cisionnel. D'une
part, il est \`{a} l'origine des donn\'{e}es qu'il veut \'{e}tudier dans la mesure o\`{u} il
interroge le m\'{e}diateur. D'autre part, il d\'{e}finit les mesures et les dimensions
pour la construction de son contexte d'analyse. Notre m\'{e}thode se charge alors
de lui fournir automatiquement une repr\'{e}sentation int\'{e}ressante en arrangeant
les modalit\'{e}s des dimensions qu'il choisit d'observer.

\bibliographystyle{RNTIBiblio}
\bibliography{eda05_bib}

\begin{thebibliography}{}

\bibitem[\protect\citeauthoryear{Barbar{\'a} et Sullivan}{1997}]{BaSu97}
Daniel Barbar{\'a} et Mark Sullivan.
\newblock {Quasi-Cubes: Exploiting Approximations in Multidimensional
  Databases}.
\newblock {\em {SIGMOD Record}}, 26(3):12--17, 1997.

\bibitem[\protect\citeauthoryear{Benzécri}{1969}]{Ben69}
Jean~Paul Benzécri.
\newblock Statistical analysis as a tool to make patterns emerge from data.
\newblock In ed.) Academic~Press (S.~Watanabe, editor, {\em Methodologies of
  Pattern Recognition}, pages 35--60, New York, 1969.

\bibitem[\protect\citeauthoryear{Bertin}{1981}]{Ber81}
Jacques Bertin.
\newblock {\em {Graphics and Graphic Information Processing}}.
\newblock de Gruyter, New York, 1981.

\bibitem[\protect\citeauthoryear{Beyer et Ramakrishnan}{1999}]{BeRa99}
Kevin Beyer et Raghu Ramakrishnan.
\newblock {Bottom-Up Computation of Sparse and Iceberg CUBEs}.
\newblock In {\em {Proceedings of ACM SIGMOD Record}}, pages 359--370, 1999.

\bibitem[\protect\citeauthoryear{Chauchat et Risson}{1998}]{ChRi98}
Jean~Hugues Chauchat et Alban Risson.
\newblock {\em BERTIN's Graphics and Multidimensional Data Analysis}, pages
  37--45.
\newblock Visualization of Categorical Data. Academic Press., 1998.

\bibitem[\protect\citeauthoryear{Choong \bgroup \em et al.\egroup
  }{2003}]{CLM03}
Yeow~Wei Choong, Dominique Laurent, et Patrick Marcel.
\newblock {Computing Appropriate Representations for Multidimensional Data}.
\newblock {\em {Data \& knowledge Engineering Journal}}, 45(2):181--203, 2003.

\bibitem[\protect\citeauthoryear{Choong \bgroup \em et al.\egroup
  }{2004}]{CLL04}
Yeow~Wei Choong, Anne Laurent, Dominique Laurent, et Pierre Maussion.
\newblock {Résumé de cube de données multidimensionnelles à l'aide de règles
  floues}.
\newblock In Revue des Nouvelles Technologies~de l'Information, editor, {\em
  4\`{e}mes Journ\'{e}es Francophones d'Extraction et de Gestion des
  Connaissances (EGC 04)}, volume~1, pages 95--106, Clermont-Ferrand, France,
  Janvier 2004.

\bibitem[\protect\citeauthoryear{Feng \bgroup \em et al.\egroup
  }{2004a}]{FFD04}
Jianlin Feng, Qiong Fang, et Hulin Ding.
\newblock {PrefixCube: Prefix-sharing Condensed Data Cube}.
\newblock In {\em {Proceedings of the 7th ACM international workshop on Data
  warehousing and OLAP (DOLAP 04)}}, pages 38--47, Washington D.C., U.S.A.,
  November 2004.

\bibitem[\protect\citeauthoryear{Feng \bgroup \em et al.\egroup
  }{2004b}]{FAE03}
Ying Feng, Divyakant Agrawal, Amr~El Abbadi, et Ahmed Metwally.
\newblock {Range CUBE: Efficient Cube Computation by Exploiting Data
  Correlation}.
\newblock In {\em Proceedings of the 20th International Conference on Data
  Engineering}, pages 658--670, 2004.

\bibitem[\protect\citeauthoryear{Inmon}{1996}]{Inm96}
W.~H. Inmon.
\newblock {\em {Building the Data Warehouse}}.
\newblock John Wiley \& Sons, 1996.

\bibitem[\protect\citeauthoryear{Kimball}{1996}]{Kim96}
Ralph Kimball.
\newblock {\em {The Data Warehouse toolkit}}.
\newblock John Wiley \& Sons, 1996.

\bibitem[\protect\citeauthoryear{Lakashmanan \bgroup \em et al.\egroup
  }{2002}]{LPH02}
Laks~V.S. Lakashmanan, Jian Pei, et Jiawei Han.
\newblock {Quotient Cube: How to Summarize the Semantics of a Data Cube}.
\newblock In {\em {Proceedings of International Conference of Very Large Data
  Bases, {VLDB'02}}}, 2002.

\bibitem[\protect\citeauthoryear{Lakshmanan \bgroup \em et al.\egroup
  }{2003}]{LPZ03}
Laks~V.S. Lakshmanan, Jian Pei, et Yan Zhao.
\newblock {QC-Trees: An Efficient Summary Structure for Semantic OLAP}.
\newblock In ACM Press, editor, {\em {Proceedings of the 2003 ACM SIGMOD
  International Conference on Management of Data}}, pages 64--75, 2003.

\bibitem[\protect\citeauthoryear{Lebart \bgroup \em et al.\egroup
  }{2000}]{LAP00}
Ludovic Lebart, Alain Morineau, et Marie Piron.
\newblock {\em {Statistique exploratoire multidimensionnelle}}.
\newblock Dunold, Paris, 3$^{e}$ édition edition, 2000.

\bibitem[\protect\citeauthoryear{Li \bgroup \em et al.\egroup }{2004}]{LHG04}
Xiaolei Li, Jiawei Han, et Hector Gonzalez.
\newblock {High-Dimensional OLAP: A Minimal Cubing Approach}.
\newblock In {\em {Proceedings of the 30th International Conference on Very
  Large Data Bases (VLDB 2004)}}, pages 528--539, August 2004.

\bibitem[\protect\citeauthoryear{Messaoud \bgroup \em et al.\egroup
  }{2005}]{BRB05}
Riadh~Ben Messaoud, Sabine Rabaseda, et Omar Boussaid.
\newblock L'analyse factorielle pour la construction de cubes de données
  complexes.
\newblock In {\em 2ème atelier Fouille de Données Complexes dans un processus
  d'extraction des connaissances, EGC 05, Paris}, pages 53--56, Janvier 2005.

\bibitem[\protect\citeauthoryear{Niemi \bgroup \em et al.\egroup
  }{2003}]{NNT03}
Tapio Niemi, Jyrki Nummenmaa, et Peter Thanisch.
\newblock {Normalising OLAP cubes for controlling sparsity}.
\newblock {\em {Data \& Knowledge Engineering }}, 46:317--343, 2003.

\bibitem[\protect\citeauthoryear{Ross et Srivastava}{1997}]{RoSr97}
Kenneth~A. Ross et Divesh Srivastava.
\newblock {Fast Computation of Sparse Datacubes}.
\newblock In {\em {Proceedings of the 23rd International Conference of Very
  Large Data Bases, {VLDB'97}}}, pages 116--125. Morgan Kaufmann, 1997.

\bibitem[\protect\citeauthoryear{Shanmugasundaram \bgroup \em et al.\egroup
  }{1999}]{SFB99}
Jayavel Shanmugasundaram, Usama~M. Fayyad, et Paul~S. Bradley.
\newblock {Compressed Data Cubes for OLAP Aggregate Query Approximation on
  Continuous Dimensions}.
\newblock In {\em {Proceedings of the fifth ACM SIGKDD International Conference
  on Knowledge Discovery and Data Mining}}, pages 223--232, August 1999.

\bibitem[\protect\citeauthoryear{Sismanis \bgroup \em et al.\egroup
  }{2002}]{SDR02}
Yannis Sismanis, Antonios Deligiannakis, Nick Roussopoulos, et Yannis Kotidis.
\newblock {Dwarf: Shrinking the PetaCube}.
\newblock In {\em Proceedings of the 2002 ACM SIGMOD international conference
  on Management of data}, pages 464--475. ACM Press, 2002.

\bibitem[\protect\citeauthoryear{Vitter \bgroup \em et al.\egroup
  }{1998}]{VWI98}
Jeffrey~Scott Vitter, Min Wang, et Bala Iyer.
\newblock Data cube approximation and histograms via wavelets.
\newblock In {\em Proceedings of the 7th {ACM} International Conferences on
  Information and Knowledge Management (CIKM'98)}, pages 96--104, Washington
  D.C., U.S.A., November 1998. Association for Computer Machinery.

\bibitem[\protect\citeauthoryear{Vitter et Wang}{1999}]{ViWa99}
Jeffrey~Scott Vitter et Min Wang.
\newblock {Approximate Computation of Multidimensional Aggregates of Sparse
  Data Using Wavelets}.
\newblock In {\em {Proceedings of the 1999 ACM SIGMOD international conference
  on Management of Data}}, pages 193--204, {Philadelphia, Pennsylvania,
  U.S.A.}, June 1999. {ACM Press}.

\bibitem[\protect\citeauthoryear{Wang \bgroup \em et al.\egroup }{2002}]{WFL02}
Wei Wang, Hongjun Lu, Jianlin Feng, et Jeffrey~Xu Yu.
\newblock {Condensed Cube: An Effective Approach to Reducing Data Cube Size}.
\newblock In {\em {Proceedings of the 18th IEEE International Conference on
  Data Engineering (ICDE'02)}}, 2002.

\end{thebibliography}

\section*{Summary}
In decision-support systems, the visual component is important for On Line
Analysis Processing (OLAP). In this paper, we propose a new approach that faces
the visualization problem due to data sparsity. We use the results of a
Multiple Correspondence Analysis (MCA) to reduce the negative effect of
sparsity by organizing differently data cube cells. Our approach does not
reduce sparsity, however it tries to build relevant representation spaces where
facts are efficiently gathered. In order to evaluate our approach, we propose
an homogeneity criterion based on geometric neighborhood of cells. The obtained
experimental results have shown the efficiency of our method.
\end{document}